\documentclass[mathleft]{an}

\usepackage{graphicx}
\usepackage{amsmath,amsfonts,amssymb}
\usepackage{times}

\def\gsim{\lower.4ex\hbox{$\;\buildrel >\over{\scriptstyle\sim}\;$}} 
\def\lsim{\lower.4ex\hbox{$\;\buildrel <\over{\scriptstyle\sim}\;$}} 

\def\[{\begin{eqnarray}}
\def\]{\end{eqnarray}}

\def\subb{\begin{subequations} \[ }
\def\sube{\] \end{subequations} }

\def\gab{\begin{gather}}
\def\gae{\end{gather}}

\newcommand{\nablasq}{\nabla^2}

\newcommand{\rot}[1]{\nabla \times #1}
\renewcommand{\rot}[1]{\mathop{\rm rot}\nolimits #1}
\renewcommand{\div}[1]{\nabla \cdot #1}
\renewcommand{\div}[1]{\mathop{\rm div}\nolimits #1}

\newcommand{\Rey}{\mathrm{Re}}

\newcommand{\Omin}{\Omega_{\mathrm{in}}}
\newcommand{\Omout}{\Omega_{\mathrm{out}}}
\def\Omend{\Omega_{\mathrm{end}}}
\def\Omtop{\Omega_{\mathrm{top}}}
\def\Ombot{\Omega_{\mathrm{bot}}}
\def\Omegab{\Omega_\mathrm{0}}

\newcommand{\Rin}{R_\mathrm{in}}
\newcommand{\Rout}{R_\mathrm{out}}

\newcommand{\muh}{\hat{\mu}}
\newcommand{\etah}{\hat{\eta}}
\def\Pm{\mathrm{Pm}}
\def\Ha{\mathrm{Ha}}
\def\mperm{\mu_0}
\def\mdiff{\eta}

\newcommand{\ord}[1]{$O(#1)$}

\newcommand{\cmnt}[1]{}
\newcommand{\comm}[1]{}
\newcommand{\ignore}[1]{}

\def\unitp{\hat{\mathbf{e}}_\phi}
\def\unitz{\hat{\mathbf{e}}_z}

\def\Omendmin{\Omega_{\mathrm{endmin}}}

\def\rring{w_{1}}
\def\rrings{w_{2}}

\def\apj{ApJ}
\def\apjl{ApJL}

\def\mnras{MNRAS}

\newcommand{\citep}[1]{\cite{#1}}
\newcommand{\citet}[1]{\cite{#1}}

 \renewcommand{\citep}[1]{}
 \renewcommand{\citet}[1]{}

\def\FIGDIR{fig}

\newcommand{\figname}[1]{\FIGDIR/#1}     
\newcommand{\bwcfigname}[1]{\FIGDIR/#1}  

\begin{document}

\setlength{\mathindent}{0cm}
\sloppy


 \Pagespan{499}{506}
 \Yearpublication{2007}%
 \Yearsubmission{2007}%
 \Month{1}%
 \Volume{328}%
 \Issue{6}%
 \DOI{10.1002/asna.200710774}%

 \title{Reduction of boundary effects in the spiral MRI experiment PROMISE}
 \author{J. Szklarski\thanks{Corresponding author: jszklarski@aip.de} }

 \institute{
 Astrophysikalisches Institut Potsdam, An der Sternwarte 16, 
 D-14482 Potsdam, Germany} 
 \received{2007 Apr 18}
 \accepted{2007 Apr 22}
 \publonline{2007 Jun 18}
 \abstract{Magnetorotational instability (MRI) is one of the most
    important and most common instabilities
    in astrophysics. It is widely accepted that it serves as a source
    of turbulent viscosity in accretion disks -- the most energy
    efficient objects in the Universe.  However it is very difficult
    to bring this process down on earth and model it in a laboratory
    experiment. Several different approaches have been proposed, one
    of the most recent is PROMISE (Potsdam-ROssendorf Magnetorotational
    InStability Experiment). It consists of a flow of a liquid metal between
    two rotating cylinders under applied current-free spiral magnetic
    field. The cylinders must be covered with plates which introduce
    additional end-effects which alter the flow and make it more difficult
    to clearly distinguish between MRI stable and unstable state. In
    this paper we propose simple and inexpensive improvement to the
    PROMISE experiment which would reduce those undesirable effects.  }


 \maketitle

\section{Introduction}

 Velikhov (1959) showed that, for ideal magnetohydrodynamics, an axial
 magnetic field applied to a flow of a liquid metal between two concentric,
 differentially rotating cylinders (Taylor-Couette flow) can lower
 the critical rotation ratio or even destabilize the flow, although it
 is hydrodynamically stable (when the Prandtl number is large enough).
 That type of instability is called magnetorotational instability,
 and not long ago Balbus \& Hawley (1991) demonstrated that it plays an
 important role in astrophysics. MRI serves as an essential mechanism
 for transporting angular momentum in a wide range of astrophysical
 objects, stellar interiors, jets and in particular it is crucial for
 process of accretion where it provides necessary amount of turbulent
 viscosity. Recent experiments by  Ji et al. (2006) suggested that purely
 hydrodynamical quasi-Keplerian flows, i.e. Taylor-Couette flows which
 resemble Keplerian disks, do not provide viscosity required to transport
 the angular momentum effectively, and therefore this instability is
 ruled out as a source of viscous turbulence in the disks.

 One of the most convenient laboratory models for MRI is still the
 magnetohydrodynamical, cylindrical Taylor-Couette flow with an imposed
 external magnetic field along the axis (R{\"u}diger \& Zhang 2001;
 Ji, Goodman \& Kageyama 2001).  For magnetic Prandtl numbers $\Pm \approx\!$ 1--10 the
 axial field (when not too strong) reduces the critical Reynolds number for
 instability and, more importantly, also introduces instability for flows
 which are always stable for nonmagnetic cases (see e.g. R{\"u}diger, Schultz \& Shalybkov
  2003; Barenghi et al. 2004).  The unstable state is characterized by
 classical Taylor vortices.  Regrettably, laboratory liquid metals possess
 very small magnetic Prandtl numbers (due to low conductivity) so that for a
 purely axial field the critical Reynolds number is of order \ord{10^6},
 and therefore vast rotation rates are necessary for MRI to grow.

 Recently it has been shown by Hollerbach \& R{\"u}diger (2005) and
 R{\"u}diger et al. (2005) that a current-free external azimuthal
 magnetic field in addition to the axial one can reduce the critical
 Reynolds number  to \ord{10^3} which makes it much easier to
 design an MRI experiment. Moreover due to symmetry-breaking there exists
 a drift of Taylor vortices associated with the configuration of the applied
 magnetic field.  The frequency of the traveling wave that can be measured 
 and compared with theory is an important feature of this type of 
 instability.

 The idea of an additional toroidal field was successfully implemented in the
 PROMISE experiment by Stefani et al. (2006) where modes corresponding
 to so called spiral (or helical) MRI  were observed for the first time
 (see also R{\"u}diger et al. 2006; Stefani et al. 2007).  Results of
 this experiment  also show that in the basic stable state, without
 any toroidal field, there exists a nonzero axial velocity field which
 arises due to presence of the rigid endplates enclosing the
 cylinders.  These plates, undoubtedly present in any real experiment,
 are responsible for additional effects which do not take place for an
 idealized infinitely long container.

 The boundary layer which exists in the vicinity of the endplates consists
 of an Ekman layer which is the result of the rotation of a rigid surface,
 and a Hartmann layer which develops when a conducting fluid is used
 and an external axial magnetic field is applied (see e.g. Ekman 1905;
 Roberts 1967).  Consequently, the global properties of the flow change when
 compared to the idealized case of infinitely long cylinders:  a secondary
 flow, i.e. two large Ekman vortices appear,  and the Hartmann current is drawn
 into the bulk of the fluid. All these effects depend on the mechanical and
 magnetic properties of the lids. In the PROMISE experiment one of the
 lids is made of copper and is attached to the outer cylinder, the other
 one is a stationary plexiglass plate.

 In this work we review simple improvements which can reduce undesired
 effects induced by the lids and provide therefore the possibility to 
 distinguish more clearly between stable and unstable states of MRI.

\section{Numerical model}

 We consider two concentric cylinders with radii \hbox{$\Rin=4$} cm, \hbox{$\Rout=8$} cm
 and height \hbox{$H=40$} cm which rotate with angular velocities $\Omin,\ \Omout$. 
 The rotation ratio $\muh=\Omout/\Omin$ is chosen in such a
 way that the flow is hydrodynamically stable, i.e. Rayleigh stability
 criterion $\partial_r(r^2\Omega)^2 > 0$, $r$ being the distance from the axis
 of rotation, is fulfilled. Through this paper we use $\muh=0.27$. For
 infinite cylinders the basic rotational profile for the flow is the Couette
 solution
 \[
  \Omegab(r) = a + \frac{b}{r^2},
 \]
 where $a,\ b$ are 
 constants dependent on radii and rotation speeds
 \[
   a=\Omin \frac{\muh-\etah^2}{1-\etah^2}, \quad\ b=\frac{1-\muh}{1-\etah^2}\Rin^2 \Omin.
 \]
 
 The external magnetic field (steady, current free) has the form of
 \[
   \vec{B_0} = B_0 \left (\frac{\beta \Rin}{r} \unitp +  \unitz \right),
 \]
 and its strength is measured by the Hartmann number 
 \[
 \Ha = B_0 \sqrt{\frac{\Rin (\Rout-\Rin)}{\mperm \rho \nu \mdiff}}.
 \]
 The magnetic properties of the conducting fluid are described by the magnetic Prandtl number which 
 is the ratio of the kinematic viscosity $\nu$ to the magnetic diffusivity $\mdiff$, $\Pm = \nu/\mdiff$, 
 $\mperm$ denotes the magnetic permeability, and $\rho$ denotes the density. 
 The Reynolds number
 $\Rey$ is defined as $\Rey=\nu^{-1} \Omin \sqrt{\Rin (\Rout-\Rin)}$.
 The liquid 
 used in the PROMISE experiment is the euticetic alloy GaInSn 
 giving $\Pm=1.4 \times 10^{-6}$.
 Thus it is reasonable to solve the MHD equations (dimensionless) in the small Prandtl number 
 limit (Youd \& Barenghi 2006; Roberts 1967; Zikanov \& Thess 1998), 
  \[
  \label{eqn-quasimhd}
  \lefteqn{ \partial_t \vec{u} + (\vec{u} \cdot \nabla)\vec{u} =
       -\nabla p + \nablasq \vec{u} + 
       \Ha^2 \rot{\vec{b}} \times \frac{\vec{B_0}}{B_0}, \label{eqn-momentumB} }  \\
  \lefteqn{ \nablasq \vec{b} = - \rot{(\vec{u} \times \vec{B_0}/B_0)}, }  \label{eqn-inducB}
   \]
 and $\div{\vec{u}}= 0,\div{\vec{b}}= 0$, where $\vec{u}$ is the velocity 
 field and $\vec{b}$ is the perturbed magnetic field. 

 We simulate the above nonlinear
 equations  for a 2\,D axisymmetric flow in cylindrical coordinates $(r, \phi, z)$. 
 For details on the numerical method and the boundary conditions see Youd \&
 Barenghi (2006) and  Szklarski \& R{\"u}diger (2006).  The cylinders are
 assumed to be perfectly  conducting, and the endplates are either conducting
 or insulating. For the latter the pseudo-vacuum approximation is used\footnote{
 However, even for copper the assumption for perfect conductors may be
 not very realistic.}.

\section{Discussion of the results} 

From the point of view of the MRI experiment we are interested in obtaining
a stable, uniform rotation profile which for subcritical characteristic
parameters is as close as possible to the idealized basic state $\Omegab$.
On the other hand we expect a clear pattern of traveling vortices 
for supercritical conditions.
For infinite cylinders with an external axial magnetic field and
liquids with $\Pm$ of our interest the basic Couette profile $\Omegab$
is not altered until a critical Reynolds number of order
\ord{10^6$--$10^7} is reached (corresponding to a rotation frequency $f\approx 100$ Hz).  
For an instability due to the additional toroidal field with
$\beta=4$ we expect $\Rey$ to be of order \ord{10^3} (implying
$f \approx 0.1$ Hz), $\Ha$ of order \ord{10}, and therefore we search
for conditions for which the flow is closest to the $\Omegab$ profile
for these parameters (for details on the critical values see 
Hollerbach \& R{\"u}diger 2005; R{\"u}diger et al. 2005).

 \begin{figure}[t]
    \center
    \includegraphics[width=\columnwidth]{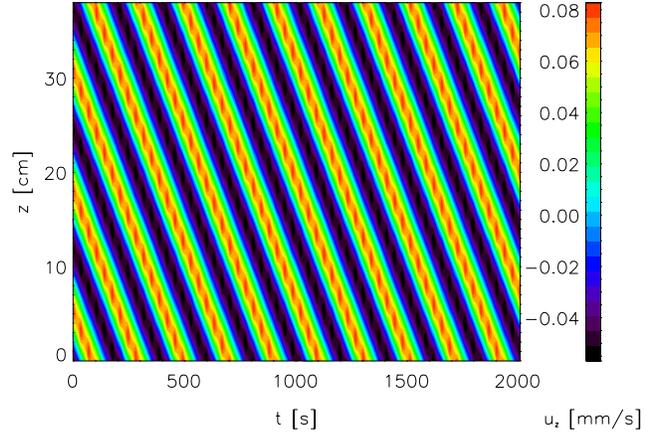}
    \caption{Profiles of \hbox{$u_z(z,t)$ at $r=\Rin + 0.6\,D$} for periodic cylinders
             just above the critical characteristic values: $\Rey=1000, \Ha=9.5, \beta=4$. 
	     The critical Reynolds number in this case is $\Rey_c=842$. 
	     } 
    \label{fig-uzsery-inf}
 \end{figure}

Figure~\ref{fig-uzsery-inf} displays values of the velocity component $u_z$
measured along the $z$ axis at $r=\Rin + 0.6D$ for supercritical values
of rotation and magnetic fields.  $\Rin=4$ cm, $\Rout=8$ cm, and the physical
properties of gallium for the viscosity and the magnetic diffusivity are used
in order to obtain values in physical scales comparable with those
of the PROMISE experiment.  The results in this figure are for cylinders with
periodic boundary conditions so that the profiles are not constrained
by end effects and are directly comparable with results from linear
theory for infinite cylinders.  We notice clear traces of the drifting
Taylor vortices.

\subsection{Reducing endplate effects}

All undesirable effects induced by the endplates arise as a consequence
of vertical shears near the boundaries. Thus we attempt to reduce
the shears by using appropriate boundary conditions.  Some experiments
(e.g. Ji et al. 2004; Noguchi et al. 2002) must deal with vast rotation
rates since a toroidal field is not applied (i.e. $\beta=0$), and the
rigidly rotating boundaries dominate the whole flow (see Hollerbach \&
Fournier 2004).  In this case it is necessary to split the end-plates
into many independently rotating rings (Kageyama et al. 2004; Burin et
al. 2006).  When the rotation rates are relatively slow, so that 
the corresponding Reynolds number is of order
\ord{10^3}, the desired result can be achieved either by allowing
the endplates to rotate rigidly and independently (see e.g. Abshagen
et al. 2004) or by splitting them into two rings which are attached to 
both cylinders. From the technical point of view the latter configuration
is easier to implement and can be considered as a possible extension
to the next spiral MRI experiment.

Firstly we consider a criterion according to which we say that the boundary
conditions are more suitable. In the basic state for subcritical parameters
and for the case of periodic cylinders, the rotational profile of the fluid is $\Omegab(r)$
and is independent of $z$, and the magnetic perturbations $\vec{b}$ are zero
everywhere.  Introducing endplates leads to the development of $z$ and
$r$ gradients in velocity, especially close to the vertical boundaries where
$\Omega(r)$ from the bulk of the fluid must match the imposed conditions at
$z=0,\ z= H$.  Consequently two Ekman vortices, new currents and magnetic
fields are generated (we assume the lids to be insulating unless
explicitly stated otherwise).  Any deviation from $\Omegab$ will result in
generating an azimuthal component of the magnetic field, $b_\phi$, which enters
the momentum equation (and in our 2D axisymmetric formulation it is
the only term which gives rise to the Lorentz force).  Vertical profiles
of $b_\phi$ in the middle, i.e. for $r=D/2$, $D=\Rout-\Rin$,  are shown in 
Fig. \ref{fig-bphi-profs} for two different boundary conditions.
If the endplates rotate rigidly with the outer cylinder, $\Omend=\Omout$,
the Ekman circulation at the bottom lid has a clockwise direction, if they
rotate with the inner cylinder, $\Omend=\Omin$, counter-clockwise -- all
the gradients have opposite sign.  We conclude that, not surprisingly,
there exists a condition with $\Omout < \Omendmin < \Omin$ for which 
the shears are minimized, and the generated magnetic field as well.

 \begin{figure}[ht]
    \center
    \includegraphics[width=\columnwidth]{\figname{bphi-profs.eps}}
    \caption{Vertical profiles for induced $b_\phi$ in the middle of the gap, $r=D/2$ 
             for $\beta=0, \Rey=100, \Ha=1$ and pseudo-vacuum boundary
	     conditions. \textemdash\textemdash~$\Omend=\Omout$, 
	     \textendash~\textendash~\textendash~$\Omend=\Omin$. 
	     } 
    \label{fig-bphi-profs}
 \end{figure}

We are interested in obtaining a rotational profile 
for which the energy in $b_\phi$ is minimized,
\[
 \label{eq:enerb}
 E_b = \iint b_\phi(r,z)^2 \mathrm{d}r \mathrm{d}z,  
\]
where the integration is done over the total volume.  As a measure of the deviation
from $\Omegab$ one could also consider, for example, the kinetic energy
of the flow, but our aim is to obtain good  rotational profiles also for $\Ha$ of
order 10 (and $\beta=0$). This is not necessarily a good approach since
the axial field can inhibit the flow velocity while the rotational profile will still
be significantly different from $\Omegab$.

 \begin{figure}[ht]
    \center
    \includegraphics[width=8.5cm]{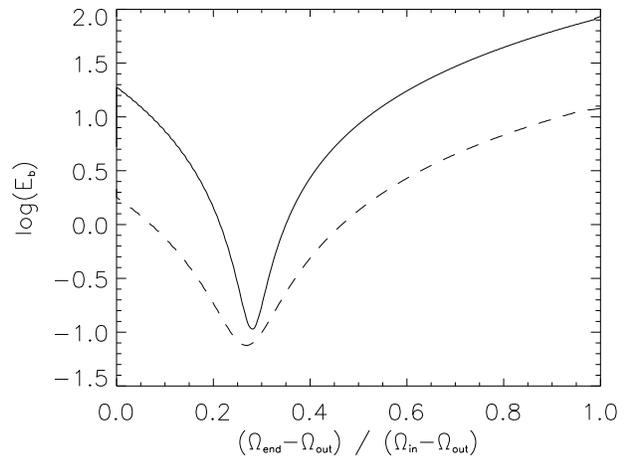}
    \caption{The magnetic energy $E_b$  as a function of angular velocity of independently rotating lid $\Omend$ for
	     $\beta\!~=~\!0, \Rey\!~=~\!100$. \textemdash\textemdash~$\Ha\!~=~\!1$, 
	     \textendash~\textendash~\textendash~$\Ha\!~=~\!10$. 
	     } 
    \label{fig-minimal-lids}
 \end{figure}

Figure \ref{fig-minimal-lids} shows how $E_b$ depends on the rotation rates of the rigid
endplates, $\Omin < \Omend < \Omout$.  We notice that a minimum occurs for
$\Omendmin \approx 0.3(\Omin - \Omout)+\Omout$ which is even three orders
of magnitude smaller than for $\Omend=\Omin$.

 \begin{figure}[ht]
    \center
    \includegraphics[width=8.5cm]{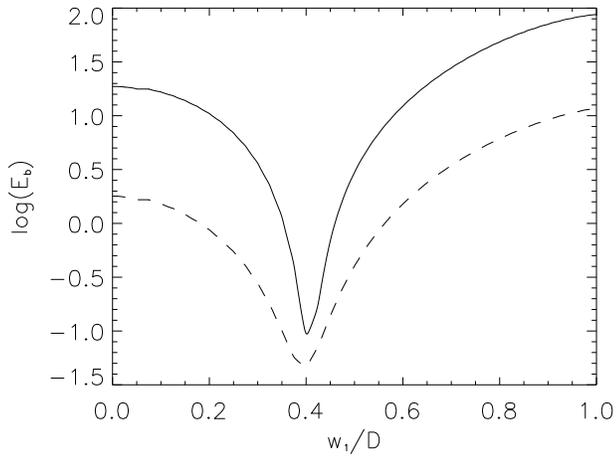}
    \caption{The magnetic energy $E_b$ as a
             function of radius of the inner ring for $\beta\!~=~\!0, \Rey\!~=~\!100$. \textemdash\textemdash~$\Ha\!~=~\!1$, 
	     \textendash~\textendash~\textendash~$\Ha\!~=\!~10$. 
	     } 
    \label{fig-minimal-rings}
 \end{figure}

When considering endplates divided into rings, we assume that the ring
attached to the inner cylinder has a width $\rring$, the other one 
attached to the outer cylinder has $\rrings = D-\rring$.  Because it is not obvious
which value for $\rring$ should be chosen we search for the optimal
$\rring$, i.e. for which $E_b$ has a minimum, by performing simulations for
several different values (Wendt 1933, for example, used 
$\rring/D = 0.5$).  From Fig.~\ref{fig-minimal-rings} we see that
the energy of the induced $b_\phi$ has a minimum for $\rring/D \approx 0.4$
which is roughly independent of the applied axial magnetic field.  It has also
been checked that the minimum holds for larger Reynolds numbers (for the
Fig.~\ref{fig-minimal-lids} as well).  Again we notice the improvement of
$E_b$ of two to three orders of magnitude when compared to one end-ring
attached either to the inner ($\rring=D$) or outer ($\rring=0$) cylinder. 
The minimum value is very similar to that for independently
rotating endplates.

 \begin{figure}[ht]
    \center
    \includegraphics[width=8.5cm]{\figname{omprofs-coudevocou-2.eps}}
    \caption{Deviations of the averaged 
    $\bar{\Omega}(r)$ from the basic state $\Omegab(r)$ for different vertical
    boundary conditions; rigidly rotating endplates (both with $\Omend$): 
    \textemdash\textemdash~$\Omend = \Omout$, $\cdots$~$\Omend=\Omin$, 
    \textendash~\textendash~\textendash~$\Omend = \Omendmin$; 
    divided into two rings: $\cdot$~\textendash~$\cdot$~\textendash~$\rring=0.5$, 
    $\cdots$~\textemdash~$\cdots$~$\rring=0.4$. 
    (a)~$\Rey=1000,~\Ha=0$, (b)~$\Rey=1000,~\Ha=10$. 
    }
    \label{fig-omprofs-coudev}
 \end{figure}
\clubpenalty=9999
\widowpenalty=9999

A qualitative view of the resulting rotational profiles gives
Fig.~\ref{fig-omprofs-coudev} which displays deviations of
$\bar{\Omega}(r)$ -- the angular velocity averaged in the $z$ domain -- 
from  $\Omegab(r)$ for different rotational properties of  
the endplates and for varied $\Rey$ and magnetic fields. The case
with independently rotating endplates refers to boundary conditions
where both lids rotate with angular velocity $\Omend$ corresponding to
the minimum value of $E_b$.  For comparison we also present the case for two
rings attached to the cylinders with equal width $\rring=\rrings=D/2$.

We see that applying independently rotating or split endplates
produce significantly more suitable profiles -- flatter and closer to 1.
We also notice that using $\rring=0.4D$ gives somewhat better 
results than $\rring=0.5D$, especially for $r>\Rin+D/2$ where the former profile
is almost flat. 

\subsection{Influence of the toroidal field}

\begin{figure}
    \includegraphics[width=17cm]{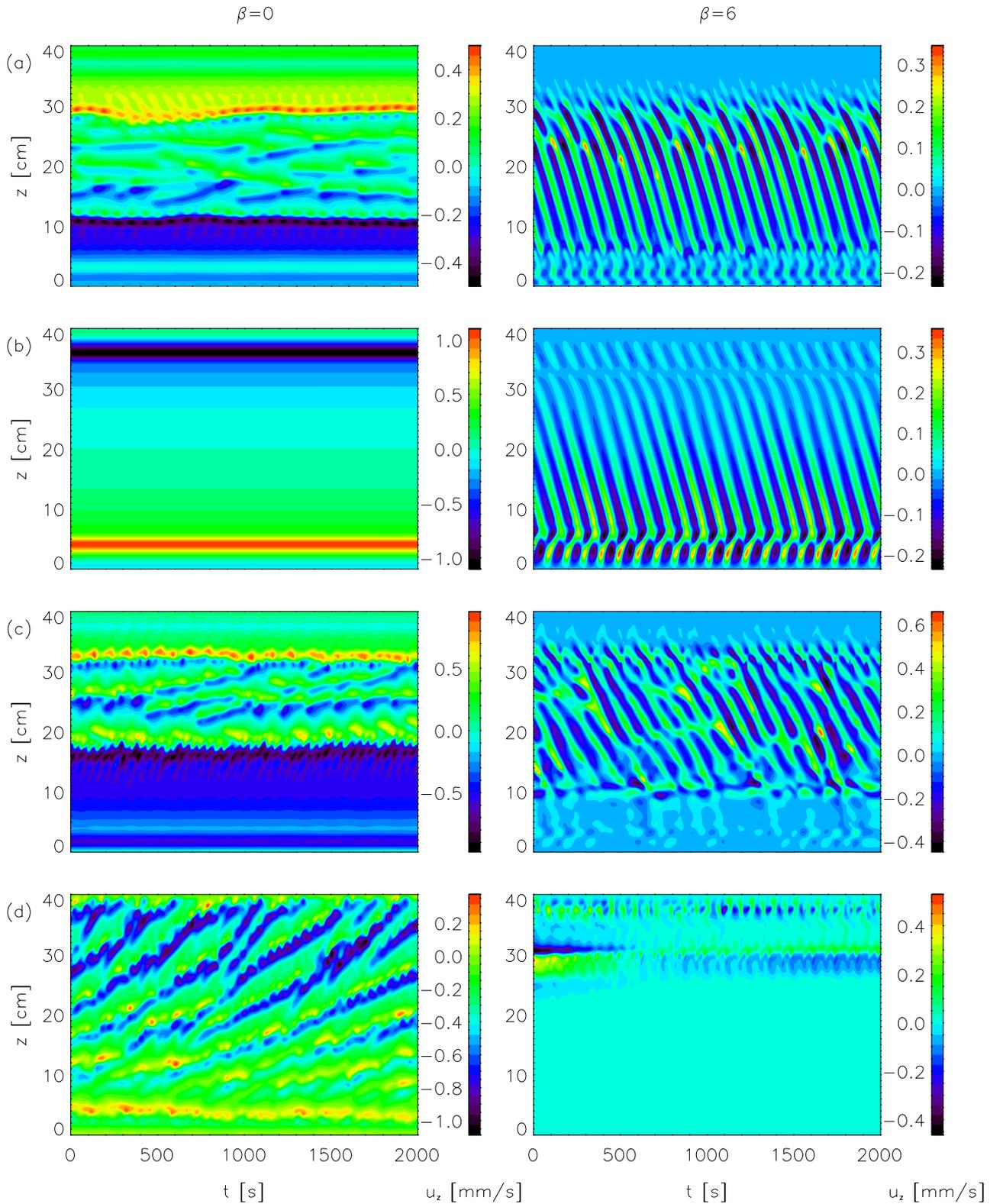}
    \caption{The axial velocity $u_z(z,t)$ at $r=\Rin + 0.6D$ as a function of time $t$ and $z$.
             \emph{Left}:
            basic state $\beta=0$. \emph{Right}:
	    $\beta=6$, the averaged $u_z(t)$ is
	    subtracted in order to eliminate the background from the 
	    velocity field, except in (d). (a): both endplates rotate rigidly with
	    $\Omend=\Omout$. 
	    (b): both endplates are divided into rings attached to the
	    cylinders; the inner ring has the width $0.4D$. (c), (d): the bottom
	    endplate is stationary $\Ombot=0$, the upper rotates with $\Omtop=\Omout$. (a),
	    (b), (c): insulating endplates, $\Rey=1775$ ($\Omin=0.377$ Hz), $\Ha=9.5$.
	    (d): perfectly
	    conducting endplates, $\Rey=1000,\ \Ha=10$. The traveling wave frequency for (a), (b), (c)
	    is respectively $f/\Omin = 0.0294,\ 0.0253,\ 0.0292$ whereas the linear stability 
	    analysis yields $f/\Omin = 0.0258$. 
            }
    \label{fig-uzsery}
 \end{figure}

Figure~\ref{fig-uzsery} shows values of the velocity component
$u_z$, similarly like Fig.~\ref{fig-uzsery-inf} but for finite cylinders. 
The velocity field $u_z$ in the basic state, i.e. $\beta=0$ for which
there is no instability, is $u_z=0$ everywhere when considering infinite
or periodic cylinders for $\Rey \approx 10^3, \Ha \approx 10$. For the enclosed
cylinder this is not the case.  In Fig.~\ref{fig-uzsery}a (left) we
present results for symmetrically, rigidly rotating (with $\Omout$), insulating
endplates. 
 We notice that $u_z$ is quite large and, more importantly,
time dependent (this is even more evident for $u_z$ 
closer to the inner cylinder). The right panel 
in this figure displays the same flow with the
toroidal field applied, $u_z(z)$ is averaged in time and subtracted in
order to filter out the background. We clearly see the
instability and structure of traveling vortices, the frequency of this motion
agrees with the predictions based on the linear analysis
(see e.g.  R{\"u}diger et al. 2005).

 \onecolumn
\twocolumn

As we have shown above, one can obtain a much better basic state for
the finite cylinders by dividing endplates into two rings. The results for such
conditions are presented in Fig.~\ref{fig-uzsery}b. We notice that the
background state quickly becomes entirely steady.  Naturally the Ekman 
pumping mechanism is still present in this case, and traces of two Ekman 
vortices can be seen.  The flow, however, is laminar.   For $\beta=6$ the pattern of
the traveling vortices is clearly more regular (cf. Fig. \ref{fig-uzsery}b, right).

When one considers two endplates with different rotational properties,
additional velocity and current gradients in the vertical direction arise
and disturb further the flow.  It is clearly seen in Fig.~\ref{fig-uzsery}c 
that disturbances exist in the case  where
the upper endplate rotates with $\Omtop=\Omout$, and the bottom one
is fixed, $\Ombot=0$. The background flow
for $\beta=0$ is highly irregular and time-dependent, especially in the
middle part of the cylinder, the circulation close to the endplates
is roughly steady. 
Nonetheless, the external $B_\phi$ produces, again, a clear
periodic motion with a frequency corresponding to that of the helical MRI.

Using conducting boundaries instead of insulating ones leads to an increase of 
the Ekman circulation and the Hartmann current, the latter 
being drawn from the plates. This current
is significantly stronger than the current generated in the Ekman-Hartmann
layer, and therefore we expect that an experiment with conducting plates would
undergo additional problems due to magnetic forces acting on the fluid.
Let us consider a perfectly conducting endplates with asymmetric
rotation (again at the top $\Omtop=\Omout$ and at the bottom $\Ombot=0$),
then there exists an important gradient in the radial current which, acting
in concert with the axial magnetic field, is strong enough to ``drag''
vortices in the direction of decreasing field strength. This situation is shown
in Fig.~\ref{fig-uzsery}d where we see a periodic vertical
motion even if $\beta=0$. Moreover, if we introduce a toroidal field
with appropriate sign (i.e. positive in this case) it will act against
the force due to the current gradient and can reduce the periodic vertical
motions in the flow (Fig.~\ref{fig-uzsery}d, right panel). If  $B_\phi$
would have a different sign both effects would interact resulting in a
highly irregular time-dependent behavior.

We  notice that in the real PROMISE experiment the
bottom endplate rotating with $\Omout$ (which, after taking into account the
directions of rotation and the applied magnetic field,
corresponds to the top endplate in our simulations) 
was made of copper, and the stationary top endplate (bottom in the simulations) 
was made of plexiglass.
Therefore an additional asymmetry in the magnetic boundary
conditions was present. Although copper is a good conductor it should not
be directly compared with perfectly conducting boundaries used in the
simulations since the latter represent stronger assumptions and induce
stronger currents. However it is clear that using insulating material
on both ends would prevent additional currents from disturbing the flow.

\subsection{Critical values}

Noting that the background state for sufficiently fast rotation
and rigidly rotating endplates $\Omend=\Omout$ is not steady, it is
interesting to investigate what happens when a spiral magnetic field
with strength below the critical value is applied. One could expect that
a viscid process (like the Ekman pumping) excites fluctuations which
could then be  amplified and, due to geometry of the applied magnetic field,
drifting.

\begin{figure}[t]
    \center
    \includegraphics[width=\columnwidth]{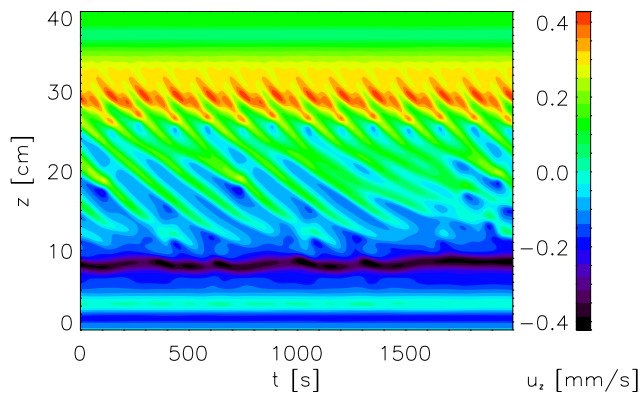}
    \caption{Profiles of $u_z(z,t)$ for $\Rey=1775,\ \Ha=9.5,\ \beta=2$ and rigidly
             rotating ends with $\Omend=\Omout$. The critical $\beta_c$ for
	     the corresponding $\Rey,\ \Ha$ in the limit of infinite cylinders  
	     is $\beta_c=2.56$, and one would expect that the traveling wave decays.
	     This is not the case for the boundary conditions shown here where a clear
	     periodic motion is visible. Its frequency $f/\Omin=0.0124$ agrees
	     with the prediction of a linear analysis for marginal stability in the limit
	     of infinite cylinders, yielding
	     $f/\Omin=0.0120$. However, the latter approach yields negative growth rate 
	     (exact numbers for frequencies and wavenumbers for the linear results presented 
	     in this paper were
	     provided by R.Hollerbach who used them to generate figures in R{\"u}diger~et~al.~(2006).
	     }
    \label{fig-uzsery-b2}
 \end{figure}

Figure~\ref{fig-uzsery-b2} shows that for endplates causing strong
disturbations the traveling wave can indeed be observed even for
subcritical characteristic values. This is also somewhat in agreement with
the experiment -- traces of moving vortices were observed for states which
are stable in the limit of infinite cylinders. We see that the amplitudes of the
vertical component of the velocity $u_z$ are almost unchanged when compared to
the background state (Fig.~\ref{fig-uzsery}a, left). Although the pattern
of the vortices is not very regular, there exists a clear frequency peak 
for the vertical traveling wave.  The frequency and the drift direction
(which is reversed by a sign change of, for example, $\beta$) corresponds
to results of the linear analysis for infinite cylinders.
This leads to the conclusion that, although excitations do not grow due to
helical MRI, still the same mechanism is responsible for the drift.

\begin{figure}[t]
    \center
    \includegraphics[width=\columnwidth]{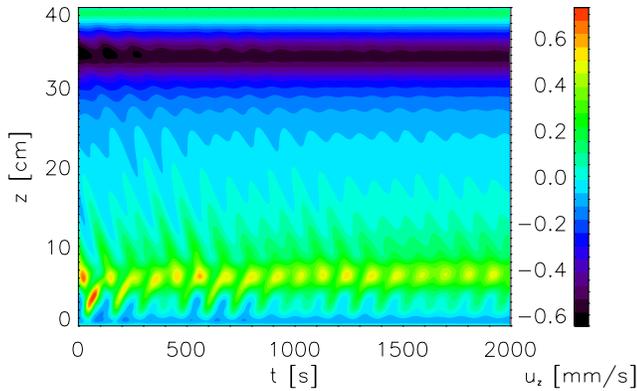}
    \caption{The velocity $u_z(z,t)$ for  $\Rey=1775,\ \Ha=9.5,\ \beta=5$ and endplates
             divided into two rings with $\rring=0.4$. Although in the limit of 
	     infinite cylinders the flow is unstable, we clearly see that
	     the disturbances (which developed after a sudden switch on of the magnetic
	     field) decay. The frequency of the decaying wave is $f/\Omin=0.0237$ which again
	     is in agreement with $f$ from the result of the linear analysis, $f/\Omin=0.0232$. 
	     } 
    \label{fig-uzsery-b5}
 \end{figure}

If two rings are used and the basic state is steady the situation
changes since the additional excitations due to the endplates are minimized.
Surprisingly, it is possible that even for supercritical parameters the
traveling wave, although excited for a moment, decays
(see Fig.~\ref{fig-uzsery-b5}). It is still possible
to get sustained instability by increasing, for example,  $\beta$
(see Fig.~\ref{fig-uzsery}b).

The reason for this damping might be height of the cylinders  
which does not match an integer value of the vertical wavenumber $k$.  
For $\Rey=1775,\ \beta=5,\ \Ha=9.5$
the corresponding wavelength is $\lambda=2\pi D/2.1728$ and does not
suit the assumed aspect ratio $\Gamma=H/D=10$.  If the height was changed to
$\Gamma = 4\lambda = 11.57$ the observed decay of the traveling wave was significantly slower,
so slow that after the sudden switching on of the external azimuthal magnetic field
the wave could be observed with the PROMISE facility still several hours later.
Bearing in mind that wavelengths for given Reynolds numbers are longer with
decreasing beta (for $\beta=3,\ k=1.45D^{-1},\ \beta=1,\ k=0.6D^{-1}$), the
constant height of the cylinders ($\Gamma=10$ in the experiment) can be an issue
when looking for critical numbers. 
It should be noted that due to the boundary layers the effective region where
the traveling wave can exists for configuration with two rings is smaller
than $\Gamma$ by approximately $0.5D$ (distances up to about $0.25D$
from the endplates are influenced by their presence).

Although endplates clearly can serve as the source of viscous excitations
and the axially traveling wave develops also for subcritical parameters,
we shall notice that there are no periodic motions in the
background state.  In this sense the ``imperfect'' background state
serves as a catalyst for the helical MRI instability.  When the endplates
are divided into rings the resulting hydrodynamic flow is laminar, and
only after the magnetic field is applied the periodic fluctuations occur,
and, moreover, their frequency corresponds exactly to that predicted from 
the linear analysis for infinite cylinders.

Liu et al. (2006) suggested that the observed fluctuations can have
theirs origin in the underlying hydrodynamical unsteady flow as reported,
for example, by Kageyama et al. (2004).  In the latter work the purely
hydrodynamic flow for $\Rey \approx 1000$ with rigidly rotating ends
$\Omend=\Omout$ and short aspect ratio $\Gamma=1$ was already unsteady.
We confirm these results with the method used here. However if
longer cylinders are used, like $\Gamma=10$, the flow becomes steady
for $\Rey=1000$, and only after imposing strong enough magnetic fields (say
$\Ha=12,\ \beta=6$) a traveling wave develops with a frequency that matches
calculations from the linear analysis.

\subsection{Differentially rotating ends}

We have also performed simulations for differentially rotating plates with
ideal Couette profile for periodic cylinders so that $\Omend(r)=\Omegab(r)$.
In another recent work (Liu, Goodman \& Ji 2007) it has been shown that for
parameters corresponding to $\Rey=1775,\ \Ha \approx 10,\ \beta \approx 4$
the traveling wave decays for such boundary conditions. We confirm this
result with our method, although our treatment of the magnetic
boundaries is simplified.

\begin{figure}[t]
    \center
    \includegraphics[width=\columnwidth]{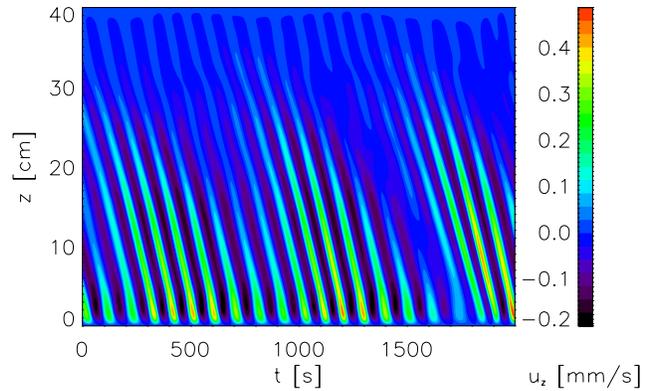}
    \caption{The velocity $u_z(z,t)$ for  $\Rey=1775,\ \Ha=9.5,\ \beta=6$ and endplates
             rotating differentially so that $\Omend=\Omegab$. 
	     } 
    \label{fig-uzsery-diffb6}
 \end{figure}

The explanation for this fact might be again the inappropriate height
of the cylinders which is far from an integer value of the expected
vertical wavelength. For these parameters $\lambda = 3.476 D$ according
to the linear theory so that less than three wavelengths can fit in the
container. On the other hand, if $\beta=6$ is used, $\lambda=2.49 D$
and then $\Gamma=10$ almost exactly corresponds to $4\lambda$. From
Fig.~\ref{fig-uzsery-diffb6} we see indeed that in this case  persistent fluctuations
exist with a frequency corresponding to the helical MRI instability. We
have also made calculations for $\beta=4$ with longer cylinders so
that $H=4\lambda D=13.90D$ and $\Gamma=5\lambda D=17.38D$. In each case a
sustained traveling wave has been observed.  It should be mentioned that
the vortices do not develop very close to the upper boundary so that it
is convenient to take longer cylinders.

\section{Summary}

It is easier to perform an experiment showing spiral MRI because a
much slower rotation of the cylinders is required for the instability
to set in compared with MRI with an axial magnetic field only.  Moreover,
there exists additional quantity, i.e. drift frequency, which is easy to
measure and can serve as an important indicator for the associated phenomena. 
It is
claimed that in the PROMISE experiment frequencies and amplitudes
corresponding to the spiral MRI were observed, and the results agreed with
theoretical calculations of both linear and nonlinear 2D simulations
(see Stefani 2007) for a review).  However it is still possible to
improve the experiment so that the basic state is a completely steady
flow.

In this paper we have presented a relatively simple and inexpensive
modification which is suitable for such an improvement. Firstly, the endplates
should be both made of insulating material and both should rotate in
the same way so that the system is symmetric in the $z$ direction. Secondly, it
is convenient to divide the lids into two rings which can be attached
to the cylinders so that no separate driving is needed. The optimal,
 width of the inner ring, in the sense of minimizing the induced 
 azimuthal magnetic field, is 1.6~cm for the current experimental setup.

Our calculations also show that spiral MRI modes can be driven by
endplate effects even for subcritical characteristic values (see
Fig.~\ref{fig-uzsery-b2}).  On the other side when providing a steady
background flow by applying rings one has to pay more attention to the 
height of the cylinders and to take into account the vertical wavelengths of the
traveling wave which depend on the magnetic configuration. For
the current aspect ratio $\Gamma=10$ and $\Rey=1775$ it is reasonable to
consider $\Ha=9.5,\ \beta=6$ which almost exactly corresponds to $\Gamma=4\lambda$.


\end{document}